\documentclass[12pt]{article}

\usepackage{graphicx}
 \newcommand{\be}{\begin{equation}}

\newcommand{\ee}{\end{equation}}
\newcommand{\al}{\alpha}
\begin{document}
\begin{titlepage}
\begin{center}
\hfill {\tt YITP-SB-05-42}\\
\hfill {\tt hep-th/0511284}\\
\vskip 20mm

{\Large {\bf Super Calabi-Yau's and Special Lagrangians}}

\vskip 10mm

{\bf  R.~Ricci \footnote{{\tt rricci@insti.physics.sunysb.edu}}}

\vskip 4mm {\em  C. N. Yang Institute for Theoretical
Physics,}\\
{\em State University of New York at Stony Brook}\\
{\em Stony Brook, NY 11794-3840, USA}\\
[2mm]

\end{center}

\vskip 1in

\begin{center} {\bf ABSTRACT }\end{center}
\begin{quotation}
\noindent We apply mirror symmetry to the super Calabi-Yau
manifold ${\bf CP}^{(n|n+1)}$ and show that the mirror can be
recast in a form which depends only on the superdimension and
which is reminiscent of a generalized conifold. We discuss its
geometrical properties in comparison to the familiar conifold
geometry. In the second part of the paper examples of
special-Lagrangian submanifolds are constructed for a class of
super Calabi-Yau's. We finally comment on their infinitesimal
deformations.

\end{quotation}

\vfill
\flushleft{November 2005}

\end{titlepage}

\eject



\section{Introduction}

Recent interest in super Calabi-Yau manifolds comes from the
duality between the topological B model on ${\bf CP}^{(3|4)}$ and
{\it perturbative} super Yang-Mills. This surprising connection
has led to a new understanding of perturbative Yang-Mills
\cite{Cachazo}. For a review see \cite{Khoze:2004ba} and
\cite{Cachazoreview}. See also
\cite{Berkovits}\cite{Neitzke}\cite{gravity}\cite{rey}\cite{PopovRB}
for a partial list of further developments. Even though this
duality can be seen as an extremely interesting counterpart of the
AdS/CFT correspondence, it has also given a new impetus to the
study of purely geometrical properties of super Calabi-Yau
manifolds. See for instance \cite{ma1}\cite{Lindstrom:2005uh} for
novel results in this direction.

Super Calabi-Yau manifolds provide an interesting arena for
studying topological strings. One remarkable conjecture is that
the topological A model on ${\bf CP}^{(3|4)}$ is equivalent to the
topological B model on a quadric inside the (super)ambi-twistor
space ${\bf CP}^{(3|3)}\times {\bf
CP}^{(3|3)}$\cite{Witten:1978xx}\cite{Neitzke}. A crucial
ingredient in this
 conjecture is mirror symmetry. The importance of supermanifolds in the context of mirror
symmetry was fully appreciated for the first time in \cite{Sethi}:
Landau-Ginzburg models which are mirror to rigid Calabi
Yau's\footnote{A Calabi-Yau is rigid when it does not have complex
structure moduli.} can be given a geometrical interpretation as
sigma models with supermanifolds as target space. The modern
language for studying mirror symmetry for toric supermanifolds has
been systematized in \cite{Aganagic:2004yh}. For other related
works see\,\cite{Kumar}\cite{ahn}\cite{belhaj}. In the first part
of the paper we will apply mirror symmetry to the super Calabi-Yau
${\bf CP}^{(n|n+1)}$ and show that the mirror can be recast in a
form which is reminiscent of a generalized conifold. The mirror
depends only on the superdimension of the supermanifold, i.e. on
the difference of bosonic and fermionic dimensions. We then
discuss its geometrical properties in comparison with the usual,
bosonic, conifold geometry.

 In Calabi-Yau compactifications special
Lagrangian submanifolds are particularly important because they
are supersymmetric cycles, known as A branes since they preserve
the A model topological charge. It is interesting to see whether
special-Lagrangian submanifolds can be constructed inside
Calabi-Yau supermanifolds. In the second part of the paper
examples of special-Lagrangians are constructed for a class of
super Calabi-Yau's in a similar spirit to what done in \cite{aga}
for local Calabi-Yau's.

Apart from those already mentioned, there other reasons of
interests in super Calabi-Yau's. The most prominent is, perhaps,
the fact that, as far as the topological A model is concerned,
certain {\it compact} bosonic Calabi-Yau's are equivalent to
(toric) super Calabi-Yau's \cite{Schwarz:1995ak}. An example is
the A model on the classic Calabi-Yau quintic in ${\bf CP}^{4}$
which is equivalent to the A model on the super-projective
Calabi-Yau space ${\bf CP}{(1,1,1,1,1|5)}$. In
\cite{aga}\cite{klemm} open string instanton corrections to the
worldvolume superpotential for some non-compact special Lagrangian
branes have been derived for a class of non-compact Calabi-Yau's
using mirror symmetry. We can then speculate that using similar
techniques, and in view of the above remarks, the study of
Lagrangian submanifolds in super Calabi Yau's could maybe help in
performing the superpotential computation in the notoriously
difficult compact Calabi-Yau case.

  The organization of the paper is as follows: We begin by
  reviewing the relevant aspects of mirror symmetry in sec.2;
  In sec.3 we apply mirror symmetry to ${\bf CP}^{(n|n+1)}$ and
discuss the mirror ``super-conifold'' geometry which arises in the
dual theory. In sec.4 we review the construction of non-compact
special Lagrangian in toric CY manifolds. This construction is
suitably extended to the supermanifold case in the next section;
In the last section we finally comment on the moduli space of
infinitesimal deformations of (super)special-Lagrangians.

\section{Gauged Linear Sigma Model and Mirror Symmetry}
In this section we review the proof of mirror symmetry for local
Calabi-Yau manifolds\cite{hori}. The proof consists in showing the
equivalence of a gauged linear sigma model and a dual
Landau-Ginzburg theory. The gauged linear sigma model reduces in
the low energy limit to a non-linear sigma model on the Calabi-Yau
manifold\cite{phase}\footnote{See also \cite{gia} for a discussion
of gauged linear sigma models on supermanifolds.}. We work in 1+1
dimensions where we study the following $(2,2)$ supersymmetric
gauge theory \be {\cal L}=\int
d^4\theta\,\left(\sum_i\bar{\Phi}^ie^{2Q_iV}\Phi^i -{1\over
2e^2}\bar{\Sigma}\Sigma\right)-{1\over 2}\int
d^2\tilde{\theta}\,t\Sigma+c.c. \label{2d}\ee The chiral fields
$\Phi^i$ have charges $Q_i$ under the U(1) gauge group with vector
superfield $V$. The twisted chiral field strength is
$\Sigma=\bar{D}_+D_{-}V$, $t=r-i\theta$ is the complexified
Fayet-Iliopoulos parameter and $d^2\tilde{\theta}$ is the twisted
chiral superspace measure $d\theta^+d\bar{\theta}^-$. In the
low-energy limit $r_0\gg 1$ the theory is equivalent to a
non-linear sigma model on the toric manifold \be
\left\{\sum_{i=1}^N Q_i|\Phi^i|^2=r_0\right\}/U(1) \ee If
$\sum_{i=1}^N Q_i=0$ the bare real F.I. parameter $r_0$ does not
renormalize. The parameter $t$ is identified with the complexified
K\"{a}hler parameter of the sigma model. The case $\sum_{i=1}^N
Q_i=0$ corresponds to a local Calabi-Yau space.

 Let us consider the``enlarged''
Lagrangian \be {\cal L}=\int d^4\theta\,\left(e^{2QV+B}-{1\over
2}(Y+\bar{Y})B\right)\label{LL1}
 \ee where $B$ is a real superfield and
 $Y$ a twisted chiral field,
$\bar{D}_+Y=D_-Y=0$, whose imaginary part has period $2\pi$.
Rewriting the superspace measure as
$d^4\theta=d\theta^+d\bar{\theta}^-\,D_-\bar{D}_+$, the field
equation for $Y$
 \be
 {\delta\over{\delta Y}} \int
d\theta^+d\bar{\theta}^-\, Y(D_-\bar{D}_+B)=0, \ee yields \be
D_-\bar{D}_+B=0\label{B1}.\ee This equation enforces the
decomposition  \be B=\psi+\bar{\psi},\label{bo} \ee where $\psi$
is a chiral superfield. Inserting this expression in (\ref{LL1})
the Lagrangian becomes

 \be {\cal
L}=\int d^4\theta\,e^{2QV+\psi+\bar{\psi}}=\int
d^4\theta\,\,\bar{\Phi}e^{2QV}\Phi\ee
 where we have introduced another chiral field $\Phi=e^{\psi}$.

 Alternatively, we can first integrate out $B$ in (\ref{LL1}) obtaining
  \be B=-2QV+\mbox{ log}\left({Y+\bar{Y}\over 2}\right).\label{B2}\ee
After inserting this result back in the Lagrangian, this yields
\be {\cal L}=\int d^4\theta\,\left(-{1\over 2}(Y+\bar{Y})\mbox{
log}(Y+\bar{Y})+QV(Y+\bar{Y})\right) \ee which, using
$\Sigma=\bar{D}_+D_{-}V$, can be rewritten as \be
 {\cal L}=\int d^4\theta\,\,\left(-{1\over
2}(Y+\bar{Y})\mbox{ log}(Y+\bar{Y})\right)+\int
d^2\tilde{\theta}Q\Sigma Y+c.c. \ee  Therefore we have shown that
the Lagrangian \be {\cal L}=\int
d^4\theta\,\left(\bar{\Phi}e^{2QV}\Phi -{1\over
2e^2}\bar{\Sigma}\Sigma\right)-{1\over 2}\int
d^2\tilde{\theta}\,t\Sigma+c.c. \label{2d}\ee
 is classically dual to
 \be
 {\cal L}=\int d^4\theta\,\left(-{1\over 2e^2}\bar{\Sigma}\Sigma-{1\over
2}(Y+\bar{Y})\mbox{ log}(Y+\bar{Y})\right)+{1\over 2}\int
d^2\tilde{\theta}\Sigma(Q Y-t)+c.c. \ee  In the duality the chiral
superfield $\Phi$ is exchanged with a {\it twisted} chiral
superfield $Y$. Comparing the different expressions (\ref{bo}) and
(\ref{B2}) for $B$ we obtain \be
\mbox{Re}Y=2\bar{\Phi}e^{2QV}\Phi. \ee In the Wess-Zumino gauge
this relation implies that the lowest components $\varphi$ and $y$
of the chiral and twisted fields satisfy
$\mbox{Re}\,y=2|\varphi|^2$. If we generalize the discussion to a
gauge theory with $n$ chiral fields $\Phi_i$, we get a dual
superpotential $\tilde{W}=\sum_i(Q_iY_i-t)\Sigma$.
 At the quantum level, non-perturbative
instanton corrections modify the dual twisted superpotential into
$\tilde{W}=\sum_i(Q_iY_i-t)\Sigma+e^{-Y_i}.$ Integrating out
$\Sigma$ gives \be \sum_i^n\,Q_iY_i=t \ee which is the dual
version of the D-term constraint of the original gauge theory.

 As an example we can consider the gauged linear sigma model with chiral
fields $(\Phi_1,\Phi_2,\Phi_3,\Phi_4)$ and charges $(1,1,-1,-1)$.
In the low-energy limit this theory is equivalent to a non-linear
sigma model on the resolved conifold ${\cal O}(-1)\oplus{\cal
O}(-1)\rightarrow {\bf CP}^1$. The lowest components of the fields
with positive charge parametrize the ${\bf CP}^1$ in the base,
while the fields with negative charge span the non-compact fibers.
The T dual-mirror theory is a Landau-Ginzburg model with dual
fields $Y_i$ that satisfy \be \mbox{Re}Y_i=|\Phi_i|^2 \ee and
superpotential $\tilde{W}=\sum_{i=1}^4e^{-Y_i}$, subject to the
constraint \be Y_1+Y_2-Y_3-Y_4=t. \ee The complex Fayet-Iliopoulos
parameter is the complexified K\"{a}hler class of the ${\bf CP}^1$
in the non linear sigma model. The Landau Ginzburg path integral
is  \be \int dY_i\,
\delta(Y_1+Y_2-Y_3-Y_4-t)\,\exp\left(\sum_{i=1}^4e^{-Y_i}\right)
\ee Solving the constraint by integrating out $Y_1$ and defining
$y_i=\exp(-Y_i)$ yields \be \int \prod_{i=2}^4{dy_i\over y_i}\,\,
\exp\left(y_2+y_3+y_4+{y_3y_4\over y_2}e^{-t}\right)\label{path}.
\ee Redefining $\tilde{y}_2=y_2/y_4,\,\tilde{y}_3=y_3/y_4$ and
introducing auxiliary variables $u,v$ in ${\bf C}$ so that \be
{1\over y_4}=\int\, dudv \,e^{\,uvy_4} \ee we can rewrite
(\ref{path}) as
\begin{eqnarray} &&\int {d\tilde{y}_2\over \tilde{y}_2}{d\tilde{y}_3\over
\tilde{y}_3}dy_4dudv\,
\exp\left(\tilde{y}_2y_4+\tilde{y}_3y_4+y_4(uv+1)+{\tilde{y}_3y_4\over
\tilde{y}_2}e^{-t}\right)\nonumber\\
&=&\int {d\tilde{y}_2\over \tilde{y}_2}{d\tilde{y}_3\over
\tilde{y}_3}dudv\,
\delta\left(\tilde{y}_2+\tilde{y}_3+uv+1+{\tilde{y}_3\over
\tilde{y}_2}e^{-t}\right),
\end{eqnarray}
where in the last step $y_4$ has been treated as a Lagrange
multiplier and integrated out. Therefore the mirror geometry, in
the patch $y_4=1$, can be regarded as the Calabi-Yau hypersurface
\be uv=\tilde{y}_2+\tilde{y}_3+{\tilde{y}_3\over
\tilde{y}_2}e^{-t}, \ee after a suitable redefinition of $u$ and
$v$. Mirror symmetry than implies that the topological A model on
the resolved conifold is equivalent to the B model on the mirror
Calabi-Yau. Note that the K\"{a}hler parameter $t$ of the initial
theory gets exchanged with the complex parameter $e^{-t}$ of the
mirror.

\section{Superconifold Geometries}

Our prototype for a supermanifold is the superprojective space
${\bf CP}^{(n|m)}$ with bosonic and fermionic coordinates
$z^i,\psi^A$ subject to the identification \be
(z^1,...,z^{n+1}|\psi^1,...,\psi^{m})\sim
\lambda(z^1,...,z^{n+1}|\psi^1,...,\psi^{m}) \ee
 where $\lambda$ is a complex number different from zero. The {\it superdimension} is
 the difference of bosonic and fermionic dimensions. In this case
 $sdim_{{\bf CP}^{(n|m)}}= n-m$.
  It is straightforward to generalize this construction to
 weighted superprojective spaces like ${\bf
CP}(Q_1,...,Q_n|P_1,...,P_m)$, where $Q_i$ and $P_i$ are
respectively the charges of the bosonic and fermionic coordinates
under the $C^\star$ action. To find a simple example of super
Calabi-Yau we may start from the supermanifold ${\bf C}^{(n+1|m)}$
with holomorphic measure $\Omega_0= dz^1\wedge \dots\wedge
dz^{n+1}\bigotimes \,\partial_{\psi^1}\dots \partial_{\psi^m}$.
The form $\Omega_0$ descends to a holomorphic form $\Omega$ on the
quotient space ${\bf CP}(Q_1,...,Q_{n+1}|P_1,...,P_m)$ if the
super Calabi-Yau condition \be
\sum_{i=1}^{n+1}Q_i-\sum_{A=1}^mP_A=0 \label{fine}\ee is
satisfied. The minus sign in front of $P_A$ is due to the fact
that $\psi$ and $\partial_\psi$ have opposite charges do the
Berezin integration rule $\int d\psi\, \psi=1$. The condition
expressed by eq.({\ref{fine}}) amounts to say that the Berezinian
line bundle of the supermanifold is trivial.

 Let us briefly review how mirror symmetry generalizes to supermanifolds.
We start with a $U(1)$ gauged linear sigma model with bosonic and
fermionic chiral fields $ \phi ^{i},\psi ^{A}$ and charges $
Q_{i},P_{A}$ respectively. The D term equation is then
\begin{equation}
\sum_{i}Q_{i}\left| \phi ^{i}\right| ^{2}+\sum_{A}P_{A}\left| \psi
^{A}\right| ^{2}=r  \label{quo}
\end{equation}
The space of vacua is the supermanifold obtained by dividing
(\ref{quo}) by the
$U(1)$ group. The dual fields which appear in the mirror theory are related to $%
\phi ^{i},\psi ^{A}$ as follows
\begin{equation}
\mbox{Re} Y^{i}=|\phi ^{i}|^{2}
\end{equation}
\begin{equation}
\mbox{Re} X^{A}=-|\psi ^{A}|^{2}
\end{equation}
This is the usual correspondence modulo the fact that $X^A$, dual
to the fermionic field $\psi^A$, picks an additional minus sign.
To guarantee that the original and the mirror supermanifolds have
the same superdimension, we need to add a couple of fermionic
fields $\eta,\chi$ to bosonic field $X$. The D term constraint
(\ref{quo}) is mirrored into
\begin{equation}
\sum_{i}Q_{i}Y^{i}-\sum_{A}P_{A}X^{A}=t\label{dmir}
\end{equation}
where $t$ is the complexified K\"{a}hler parameter. The
superpotential for the mirror Landau Ginzburg description is
similar to the bosonic case
\begin{equation}
W=\sum_{i=1}e^{-Y^{i}}+\sum_{A=1}e^{-X^{A}}\left( 1+\eta ^{A}\chi
^{A}\right) \label{LG}
\end{equation}
modulo the presence of the additional contribution
$\sum_{A=1}e^{-X^{A}}\eta ^{A}\chi ^{A}$ for the fermionic fields.
It is intended that the fields satisfy the D term
constraint(\ref{dmir}). Using this technique, it has been
shown\cite{Aganagic:2004yh} that the mirror of
 ${\bf CP}^{\left( 3|4\right) }$  is a super Calabi-Yau hypersurface
\begin{equation}
\sum_{i=1}^{3}x_{i}y_{i}+x_{i}+1+e^{t}y_{1}y_{2}y_{3}+\eta
_{i}\chi _{i}=0.\label{pista}
\end{equation}
 In the limit $t\rightarrow -\infty
$, eq.(\ref{pista}) can be thought as a quadric in a patch of
${\bf CP}^{\left( 3|3\right) }\times {\bf CP}^{\left( 3|3\right)
}$ with local inhomogeneous coordinates $\left(
x_{i},\eta_{i}\right) $ and $\left( y_i,\chi _{i}\right)$.

We now apply mirror symmetry to the supermanifold ${\bf
CP}^{(n|n+1)}.$ The path integral for the mirror Landau Ginzburg
model is
\begin{equation}
\int \prod_{i=1}^{n+1}dY_{i}dX_{i}d\eta _{i}d\chi _{i}\delta
\left( \sum_{i=1}^{n+1}Y_{i}-\sum_{i=1}^{n+1}X_{i}-t\right)
\exp\left(
\sum_{i=1}^{n+1}e^{-Y_{i}}+\sum_{i=1}^{n+1}e^{-X_{i}}\left( 1+\eta
_{i}\chi _{i}\right) \right)  \label{ori}
\end{equation}
Solving the delta function constraint by integrating out $X_{1}$
yields
\begin{eqnarray}
&&\,\,\int \prod_{i=1}^{n+1}dY_{i}d\eta _{i}d\chi
_{i}\prod_{j=2}^{n+1}dX_{j} \\
&&\exp\left(
\sum_{i=1}^{n+1}e^{-Y_{i}}+e^{t}%
\prod_{i=1}^{n+1}e^{-Y_{i}}\prod_{j=2}^{n+1}e^{X_{j}}\left( 1+\eta
_{1}\chi _{1}\right) +\sum_{i=2}^{n+1}e^{-X_{i}}\left( 1+\eta
_{i}\chi _{i}\right) \right)
\end{eqnarray}
Now we integrate over the fermionic fields $\eta _{1},\chi _{1}$
\begin{eqnarray}
&&\,\,\int
\prod_{i=1}^{n+1}dY_{i}e^{-Y_{i}}\prod_{j=2}^{n+1}dX_{j}e^{X_{j}}d\eta
_{j}d\chi
_{j} \\
&&\exp\left(
\sum_{i=1}^{n+1}e^{-Y_{i}}+e^{t}%
\prod_{i=1}^{n+1}e^{-Y_{i}}\prod_{j=2}^{n+1}e^{X_{j}}+%
\sum_{i=2}^{n+1}e^{-X_{i}}\left( 1+\eta _{i}\chi _{i}\right)
\right)
\end{eqnarray}
We did not include an irrelevant overall factor  $e^{-t}$. We
integrate in a similar way over all the remaining fermionic
coordinates except $\eta_{n+1},\chi_{n+1}$ obtaining
\begin{eqnarray*}
&&\,\,\int
\prod_{i=1}^{n+1}dY_{i}e^{-Y_{i}}\prod_{j=2}^{n+1}dX_{j}e^{X_{n+1}}d\eta
_{n+1}d\chi
_{n+1} \\
&&\exp\left(
\sum_{i=1}^{n+1}e^{-Y_{i}}+e^{t}%
\prod_{i=1}^{n+1}e^{-Y_{i}}\prod_{j=2}^{n+1}e^{X_{j}}+%
\sum_{i=2}^{n}e^{-X_{i}}+e^{-X_{n+1}} \left( 1+\eta _{n+1}\chi
_{n+1}\right) \right).
\end{eqnarray*}
The field redefinition $y_{i}=e^{-Y_{i}},x_{i}=e^{-X_{i}}$ allows
to rewrite the path integral as
\begin{eqnarray}
&&\,\,\int \prod_{i=1}^{n+1}dy_{i}\prod_{j=2}^{n}{dx_{j}\over
x_j}{dx_{n+1}\over x_{n+1}^2}d\eta _{n+1}d\chi _{n+1}\\
&&\exp\left(
\sum_{i=1}^{n+1}y_{i}+e^{t}%
\prod_{i=1}^{n+1}y_{i}\prod_{j=2}^{n+1}x_{j}^{-1}+%
\sum_{i=2}^{n}x_{i}+x_{n+1}\left( 1+\eta _{n+1}\chi _{n+1}\right)
\right).\nonumber
\end{eqnarray}
Using the rescaling $\widetilde{y}_{1}=y_{1},\widetilde{y}_{j}=y_{j}/x_{j}$, for $%
j=2,...,n+1$  we can recast the result as
\begin{eqnarray}
&&\,\,\int \prod_{i=1}^{n+1}d{\tilde
y}_{i}\prod_{j=2}^{n}dx_{j}{dx_{n+1}\over x_{n+1}}d\eta _{n+1}d\chi _{n+1}\\
&&\exp\left(
{\tilde y}_{1}+\sum_{i=2}^{n+1}{\tilde y}_{i}x_i+e^{t}%
\prod_{i=1}^{n+1}{\tilde y}_{i}+%
\sum_{i=2}^{n}x_{i}+x_{n+1}\left( 1+\eta _{n+1}\chi _{n+1}\right)
\right)\nonumber
\end{eqnarray}
By introducing the auxiliary bosonic variables $u,v$, we rewrite
the factors $1/x_{n+1}$ in the path integral measure as follows:
\begin{equation}
\frac{1}{x_{n+1}}=\int dudve^{uvx_{n+1}}
\end{equation}
The integral then becomes
\begin{eqnarray}
&&\,\,\int \prod_{i=1}^{n+1}d{\tilde
y}_{i}\prod_{j=2}^{n+1}dx_{j}d\eta _{n+1}d\chi _{n+1}dudv\\
&&\exp\left(
{\tilde y}_{1}+\sum_{i=2}^{n+1}{\tilde y}_{i}x_i+e^{t}%
\prod_{i=1}^{n+1}{\tilde y}_{i}+%
\sum_{i=2}^{n}x_{i}+x_{n+1}\left( 1+\eta _{n+1}\chi
_{n+1}+uv\right) \right)\nonumber
\end{eqnarray}that is

\begin{eqnarray}
&&\,\,\int \prod_{i=1}^{n+1}d{\tilde
y}_{i}\prod_{j=2}^{n+1}dx_{j}d\eta _{n+1}d\chi _{n+1}dudv\\
&&\exp\left(
{\tilde y}_{1}\left(1+e^{t}%
\prod_{i=2}^{n+1}{\tilde y}_{i}\right)+\sum_{i=2}^{n}x_i({\tilde y}_{i}+1)%
+x_{n+1}\left( 1+\eta _{n+1}\chi _{n+1}+uv+{\tilde y}_{n+1}\right)
\right)\nonumber
\end{eqnarray}
This form is convenient because the integrations over ${\tilde
y}_{1},x_{i=2,...,n+1}$ give delta functions
\begin{equation}
\int \prod_{i=2}^{n+1}d{\tilde y}_{i}dudv\,\delta\left( 1+\eta
_{n+1}\chi _{n+1}+uv+{\tilde
y}_{n+1}\right)\prod_{i=2}^{n}\delta({\tilde y}_{i}+1)\delta\left(1+e^{t}%
\prod_{i=2}^{n+1}{\tilde y}_{i}\right)\label{muor}
\end{equation}
Solving the last delta function constraint in eq.(\ref{muor}) we
get:
\begin{equation}\label{tile}
{\tilde y_{n+1}}=-{e^{-t}\over \prod_{i=2}^{n}{\tilde y_{i}}}.
\end{equation}
Imposing the constraints $\prod_{i=2}^{n}\delta({\tilde y}_{i}+1)$
on eq.(\ref{tile}) then yields
\begin{equation}
{\tilde y_{n+1}}=\pm e^{-t}
\end{equation} the plus and minus signs being respectively when $n$
is even or odd. We can then solve the last delta function
appearing in (\ref{muor}) obtaining
\begin{equation}
1+\eta _{n+1}\chi _{n+1}+uv\pm e^{-t}=0. \label{sconi}
\end{equation}
We have 2 bosonic variables $u,v$ with eq. (\ref{sconi}) as
constraint and two fermionic coordinates. The superdimension is
therefore -1 and matches the superdimension of ${\bf
CP}^{(n|n+1)}$. So we see that the mirror geometry (apart from the
sign difference in the $n$ even and $n$ odd cases) does not really
depend on $n$, but only on the superdimension. So we have recast
the mirror geometry in the form \be uv+\eta\chi=a
\label{sconi0}\ee in ${\bf C}^{(2|2)}$. The equation degenerates
to $uv+\eta\chi=0$ for $t=0$ and $n$ even, or $t=i\pi$ and $n$
odd. The form of equation (\ref{sconi0}) is reminiscent of the
deformed conifold equation
\begin{equation}
xy+uv=a      \label{coni}
\end{equation}
 in ${\bf C}^4$.  For this reason we will refer to equation (\ref{sconi0})
 as the ``superconifold''.

 We want now to compare the two conifold-like geometries. Let us begin reviewing some
 aspects of the geometry of the familiar conifold.  The complex deformation
parameter $a$ resolves the node singularity of the conifold
geometry $xy+uv=0$, by replacing the origin with a 3-sphere. The
deformed conifold is topologically $T^{*}S^3$, i.e. the cotangent
bundle of a $S^3$. This can be seen as follows. We start by
rewriting the defining equation as $\sum_{i=1}^{4}x_i^2=a$. The
constant can always be taken real by suitably redefining the
$x_i$'s. Decomposing $x_i$ into real and imaginary parts as
$x_i=v_i+iw_i$, we can write equivalently
\begin{equation} \sum_{i=1}^{4}v_i^2-w_i^2=a,
\,\,\ \sum_{i=1}^{4}v_i w_i=0.\label{star}
\end{equation}
Interpreting ${w_i}$ as coordinates along the fiber we see that
the base is an $S^3$ with coordinates ${v_i}$'s.  The base of the
bundle is an example of ``special Lagrangian submanifold''. A real
middle-dimensional submanifold $L$ of a K\"{a}hler manifold is
Lagrangian if the restriction of the K\"{a}hler form on $L$ is
zero. If in addition $\mbox{Im} \Omega_L=0$  also holds, the
submanifold is called special Lagrangian. Here the K\"{a}hler form
on $T^{*}S^3$ can be written as $2\sum_{i=1}^{4}dv_i dw_i$. This
is clearly zero on the base, since $w_i=0$. Similarly one can
verify that the imaginary part of the holomorphic measure is zero
when restricted to the base. Therefore the base $S^3$ is a special
Lagrangian submanifold inside the non compact Calabi-Yau $\,T^*
S^3$.

We can follow a similar analysis for $uv+\eta\chi=a$. Let us begin
by rewriting equation (\ref{sconi0}) as
\begin{equation}
u_1^2+u_2^2+\lambda_{\alpha}\lambda^{\alpha}=a, \label{star2}
\end{equation}
by identifying $\chi=\sqrt{2}\lambda^1$ and
$\eta=\sqrt{2}\lambda^2$. We use the following decompositions into
real and imaginary parts, $u_i=v_i+iw_i$ and
$\lambda_{\alpha}=\eta_{\alpha}+i\nu_{\alpha}$. Equation
(\ref{star2}) is then equivalent to
\begin{equation}
\sum_{i=1}^{2}v_i^2-w_i^2+\sum_{\alpha=1}^{2}\eta_{\alpha}\eta^{\alpha}-\nu_\alpha\nu^\alpha=a,
\,\,\ \sum_{i=1}^{2}v_i
w_i+\sum_{\alpha=1}^{2}\eta_\alpha\nu^\alpha=0.\label{star3}
\end{equation}
We interpret $(w_i,\nu_\alpha)$ as coordinates in the fiber and
$(v_i,\eta_{\alpha})$ as parameterizing the supersphere
$S^{(1|2)}$,
$\sum_{i=1}^{2}v_i^2+\sum_{\alpha=1}^{2}\eta_{\alpha}\eta^{\alpha}=a$,
in the base. Extending the notion of special Lagrangian
submanifold to supermanifolds, we can ask whether $S^{(1|2)}$ is
(super)special-Lagrangian. Formally then, we could view
$uv+\eta\chi=a$ as $T^*S^{(1|2)}$. The standard K\"{a}hler form of
${\bf C}^{(2|2)}$, when expressed in terms of $v_i,w_i,\eta,\nu$,
is\footnote{Note that the superform $d\eta$ and $d\chi$ are
commuting objects. For more about conventions on superforms I
refer to sec.5.}
$\sum_{i}du_id\bar{u}_i+\sum_{\alpha}d\lambda_\alpha
d\bar{\lambda}_\alpha=\sum_i dv_i
dw_i+\sum_{\alpha}(d\eta_\alpha)^2+(d\nu_\alpha)^2$ and does not
reduce to zero on the base $w=\eta=0$. We can nevertheless make a
``mild'' modification on the fermionic part of the K\"{a}hler form
of ${\bf C}^{(2|2)}$ such that its restriction on the
superconifold is zero. That is we consider the superconifold as
embedded in a new supermanifold ${\bf C}^{(2|2)}_\star$ with
modified K\"{a}hler form
$\omega=du_id\bar{u}_i+\epsilon_{\alpha\beta}d\lambda_{\beta}d\bar{\lambda}^{\alpha}$.
The new space is still super Calabi-Yau as one can easily verify
by checking that the super Monge-Ampere equation is satisfied. The
new K\"{a}hler form can be further reduced to
\begin{equation}
\omega=-2i\sum_{i=1}^{2}dv_i dw_i-2i\sum_{\al=1}^{2}d\eta_\alpha
d\nu^\alpha.
\end{equation} and its restriction on
$S^{(1|2)}$ is zero.  Since the imaginary part of the holomorphic
measure is also zero when restricted to the base, we can view
$S^{(1|2)}$ as a special Lagrangian submanifold.

Another well known resolution of the ordinary conifold singularity
is the so called ``small resolution'' which, in mathematical
terms, consists in replacing the conifold with the bundle
$\mathcal{O}(-1)\oplus \mathcal{O}(-1)\rightarrow {\bf CP}^1$. In
this case the origin is replaced with an $S^2$. We can give an
explicit description as follows. We replace the singular conifold
geometry $xy-uv=0$ with the following equation
\begin{equation}
\left(%
\begin{array}{cc}
  x & u \\
  v & y \\
\end{array}%
\right)\left(%
\begin{array}{c}
  z_1 \\
  z_2 \\
\end{array}%
\right)=0 \label{matrix}\end{equation} where $(z_1,z_2)\in {\bf
CP}^{1}$. Since $(z_1,z_2)$ is always different from zero, we have
\be \mbox{det}\left(%
\begin{array}{cc}
  x & u \\
  v & y \\
\end{array}%
\right)=0, \ee i.e. the conifold equation. Outside the origin of
${\bf C}^4$, eq.(\ref{matrix}) simply specifies a point in ${\bf
CP}^{1}$ and therefore the new geometry coincides with the old
one. At the origin instead, $(z_1,z_2)$ are unconstrained and
therefore we have a full ${\bf CP}^1$ which resolves the node
singularity. In the supermanifold context we can proceed similarly
considering the following ``resolution'':
\begin{equation}
\left(%
\begin{array}{cc}
  u & \eta \\
  \chi & v \\
\end{array}%
\right)\left(%
\begin{array}{c}
  z_{even} \\
  z_{odd} \\
\end{array}%
\right)=0 \label{rpo}\end{equation}where now $(z_{even}|z_{odd})$
lives in ${\bf C}^{(1|1)}/{\bf C}^*\equiv{\bf C}^{(0|1)}$.  The
super-conifold can be obtained from the Berezinian
\begin{equation}
\mbox{sdet}\left(%
\begin{array}{cc}
  u & \eta \\
  \chi & v \\
\end{array}%
\right)=0.
\end{equation}
Therefore in this case the singularity at the origin is replaced
by ${\bf C}^{(0|1)}$. Note that, using the ${\bf C}^*$ action,
$(z_{even}|z_{odd})\sim (1|\psi)$, and that $u=-\eta\psi$ and
$\chi=-v\psi$. Moreover since ${\bf C}^{(0|1)}$, differently from
${\bf CP}^{1}$ in the bosonic case, can be covered with only one
patch, the resolution (\ref{rpo}) can be globally parameterized by
$(v|\,\eta,\psi)$ and therefore coincides with ${\bf C}^{(1|2)}$.

As a final comment let us note that the familiar conifold equation
can be given a gauge invariant description in terms of four chiral
superfields $(\phi_1,\phi_2,\phi_3,\phi_4)$ with $U(1)$ charges
$(1,1,-1,-1)$. The gauge invariant combinations $x\equiv x_1x_3$,
$u\equiv x_1x_4$, $v\equiv x_2x_3$, $y\equiv x_2x_4$ satisfy, as a
constraint, the conifold equation. In the present context we would
have to modify the charge assignment to $(1,1,1,1)$ and therefore
we do not have anymore a gauge invariant description.

\section{Lagrangian Submanifolds}

We have seen an example of a (super)special Lagrangian in the
discussion of the ``superconifold'' in the last section. In the
second part of the paper we want provide further interesting
examples of special Lagrangians inside super-toric varieties and
discuss their geometric properties.

 We begin by reviewing the construction of Lagrangian submanifolds in
${\bf C}^{n}$\cite{harvey}\cite{aga}\cite{klemm}.  This
construction will be extended to supermanifolds in the next
section. We use a polar coordinate system, i.e. we parameterize
${\bf C}^{n}$ with $\{
|z^{i}|^{2},\theta ^{i}\}.$ The K\"{a}hler form for $%
{\bf C}^{n} $ is then
\begin{equation}
\omega =\sum_{i}d|z^{i}|^{2}\wedge d\theta ^{i}.
\end{equation}
A Lagrangian submanifold $L$ is a real $n$-dimensional subspace
satisfying $\omega_{|L}=0$, i.e. the restriction of the K\"{a}hler
form on $L$ is zero. An obvious Lagrangian is therefore $\theta
^{i}=$const., $\forall i$ and no constraints on the $|z^{i}|$'s.
Let us call $L_0$ this Lagrangian. More interesting Lagrangians
can be built out of this one. Inside $L_{0}$ we consider the
subspace
\begin{equation}
\sum_{i}q_{i}^{\alpha }|z^{i}|^{2}=c^{\alpha },\,\,\,\alpha
=1,...,n-r. \label{fin}
\end{equation}
This is a real $r$-dimensional subspace of $L_0.$ \ We can trade
the $n$ redundant variables $|z^{i}|$ for the coordinates
$s^{\beta },\beta =1,...,r, $ through the linear transformation
\begin{equation}
|z^{i}|^{2}=v_{\beta }^{i}s^{\beta }+d^{i},\,\,\beta =1,...,r.
\end{equation}
To satisfy eq.$\left( \ref{fin}\right) $  we need to impose
$v_{\beta }^i q^{\alpha }_i =0$ and $q_{i}^{\alpha
}d^{i}=c^{\alpha }.$ \ Since this subspace, that we call
 ${\cal L}$, is contained in $%
L _{0}$ we trivially have $\omega_{|}=0$ but it is not Lagrangian
since it is not middle-dimensional. We can nevertheless get a
Lagrangian submanifold fibering over each point of ${\cal L}$ a
torus $T^{n-r}$ by imposing that the angles $\theta ^{i}$ satisfy
\begin{equation}
\sum_{i}v_{\beta }^{i}\theta ^{i}=0.  \label{pup}
\end{equation}
It is easy then to check that $\omega_{|}=0$:
\begin{eqnarray}
\omega &=&\sum_{i}d|z^{i}|^{2}\wedge d\theta ^{i}=\sum_{i,\beta
}v_{\beta
}^{i}ds^{\beta }\wedge d\theta ^{i} \\
&=&\sum_{\beta }ds^{\beta }\wedge d\left( \sum_{i}v_{\beta
}^{i}\theta ^{i}\right).
\end{eqnarray}
Using $v_{\beta }^i q^{\alpha }_i =0$, eq.$\left( \ref
{pup}\right) $ can be satisfied by choosing $\theta
^{i}=q_{i}^{\alpha }\varphi _{\alpha }.$ The angles $\varphi
_{\alpha }$ span the torus $T^{n-r}$.

Consider now the Calabi-Yau $Y={\bf C}^{n}//G$ where
$G=U(1)^{n-k}$ and with D-term equations
\begin{equation}
\sum_{i}Q_{i}^{a}\left| z^{i}\right|
^{2}=r^{a},\,\,\,\,\,a=1,...,n-k. \label{dter}
\end{equation}
The Calabi-Yau condition amounts to requiring
$\sum_{i}Q_{i}^{a}=0,\,\forall a.$ The Lagrangian submanifolds of
${\bf C}^{n}$ descend to $Y$ if the condition $v_{\beta
}^{i}\theta ^{i}=0$ is well defined, i.e. preserved, in the
quotient. The action of the a$^{th}$ $U(1)$ group on the phase
$\theta ^{i}$ of the i$^{th}$ chiral field is $\theta
_{i}\rightarrow \theta _{i}+Q_{i}^{a}\varphi ^{a}.$\ Therefore, to
preserve $v_{\beta }^{i}\theta ^{i}=0$, we need to impose
\begin{equation}
\sum_{i}Q_{i}^{a}v_{\beta }^{i}=0.\label{imp}
\end{equation}
Let us consider some examples. \newline {\it Example 1}
\newline Consider the following locus in ${\bf C}^3$
\begin{equation}
2|z_1|^2-|z_2|^2-|z_3|^2=c \label{ex1}
\end{equation}
Using $\theta ^{i}=q_{i}^{\alpha }\varphi _{\alpha }$ gives
$\theta_1=2\phi$ and $\theta_2=\theta_3=-\phi$. In this case we
have a $S^1$ fibration, parameterized by $\phi$, over the locus
(\ref{ex1}). The vectors $v_{\beta}$ are $v_1=(1,1,1)$,
$v_2=(0,1,-1)$. \newline {\it Example 2}\newline As a second
example we take in ${\bf C}^4$
\begin{equation}
2|z_1|^2-|z_2|^2-|z_3|^2=c^1, \,\,\, \label{ex2}
|z_1|^2-|z_4|^2=c^2
\end{equation} To build a Lagrangian we fiber a torus over the base (\ref{ex2})
parameterized by the angles $\phi_1,\phi_2$. The condition $\theta
^{i}=q_{i}^{\alpha }\varphi _{\alpha }$ yields $\theta_1=2\phi_1+
\phi_2$, $\theta_2=-\phi_1$, $\theta_3=-\phi_1$ and
$\theta_4=-\phi_2$. The vectors $v_{\beta}$ are $v_1=(1,1,1,1)$
and $v_2=(0,1,-1,0)$. This Lagrangian will be preserved in the
K\"{a}hler quotient ${\bf C}^4//U(1)$ if the charges ${Q_i}$
satisfy (\ref{imp}), i.e. $Q_1+Q_2+Q_3+Q_4=0$ and $Q_2=Q_3$. Due
to the first condition the quotient is automatically a Calabi-Yau
manifold.
\newline {\it
Example 3} \newline As a final example we consider the Lagrangian
(A brane)
\begin{equation}
|z_2|^2-|z_4|^2=c^1, \,\,\, \label{ex2}
|z_3|^2-|z_4|^2=c^2\label{Ab}
\end{equation}
in the resolved conifold geometry $\mathcal{O}(-1)\oplus
\mathcal{O}(-1)\rightarrow \bf{P}^1$. As quotient of ${\bf C}^4$
this threefold is characterized by the $U(1)$ charges
$Q=(1,1,-1,-1)$.

All the examples considered so far are actually {\it special}
Lagrangian submanifolds. In this context the special Lagrangian
condition is equivalent to requiring $\sum_{i}q_{i}^{\alpha }=0$.
``A branes''in non-compact Calabi-Yau threefold like (\ref{Ab})
have been studied in depth in \cite{aga}\cite{klemm} where the
problem of counting holomorphic instantons ending on special
Lagrangian submanifolds was solved using mirror symmetry.

\section{Super Lagrangian Submanifolds}

We now want to generalize the previous construction to toric super
Calabi-Yau manifolds. The idea would be to start from constructing
examples of super Lagrangians in ${\bf C}^{(n|m)}$ and
successively study the conditions under which they descend to
super Calabi-Yau's built as quotients of ${\bf C}^{(n|m)}$. The
supermanifold ${\bf C}^{(n|m)}$ has K\"{a}hler potential
$z^i\bar{z}^i+\psi^A\bar{\psi}^A$ and
 super-K\"{a}hler form \be
d|z^i|^2\wedge d\theta^i+d\psi^Ad\bar{\psi}^A. \ee Our conventions
for (anti-)commutations relation for superforms are as follows
 \be
\omega_1\omega_2=(-1)^{a_1a_2+b_1b_2}\omega_2\omega_1 \ee where
$a_i$ and $b_i$ are respectively the superform degree and the
${\bf Z}_2$ Grassmann grading of $\omega_i$. For example $dz$ has
$a=1$ and $b=0$ while $d\psi$ has $a=b=1$. Using this rule we
obtain the familiar wedge product anticommutation rule $dz d{\bar
z}=-d{\bar z} dz$ but also in particular $d\psi d{\bar
\psi}=d{\bar \psi} d\psi$. One should not confuse the commuting
$d\psi^A$'s entering in the K\"{a}hler form with the
anti-commuting $d\psi^A\equiv\partial_{\psi^A}$'s in the
holomorphic measure. The $d$ operator is
$d=dz^i\partial_{z^i}+d\psi^A\partial_{\psi^A}$ with Leibnitz
rule\footnote{With this convention $\psi d\psi=-d\psi\, \psi$.}
$d(\omega_1\omega_2)=d\omega_1\,\omega_2+(-1)^r\omega_1d\omega_2$
if $\omega_1$ is a superform of degree $a=r$.

In ${\bf C}^{n}$ the prototype for a Lagrangian submanifold is the
real locus \be \theta^i=\theta_0^i,\,\,\,\, i=1,...,n
\label{pillo}\ee with $\theta_0^i$ constant. Since the notion of
polar coordinates does not extend to fermionic variables we need a
new way to think about eq.(\ref{pillo}).  The Lagrangian
submanifold (\ref{pillo}) can be rewritten as
$z^i=e^{2i\theta_0^i}{\bar z}^i$ and this form can be easily
generalized to the supermanifold case as follows
 \be
 z^i=e^{2i\theta_0^i}{\bar z}^i,\,\,\,\psi^A=e^{2i\Theta_0^A}{\bar
 \psi}^A.\label{superlago}
\ee This is a middle-dimensional submanifold of ${\bf C}^{(n|m)}$
but it fails to satisfy the condition $\omega_|=0$. Indeed the
fermionic part $d\psi^A d\bar{\psi}^A$ of the super-K\"{a}hler of
${\bf C}^{(n|m)}$ restricts on (\ref{superlago}) to
$e^{2i\Theta_0^A}d\psi^Ad\psi^A\neq 0$.

A real submanifold like (\ref{superlago})
 becomes Lagrangian if we modify the fermionic part of $\omega$
 and make it ``symplectic'' in the following sense:
 \be
\omega=i\sum_{i=1}^{n}dz^id\bar{z}^i+i\sum_{k=1}^{m}\epsilon_{A_k\,B_k}d\psi^{A_k}d\bar{\psi}^{B_k},
\label{kahler} \ee We will denote the corresponding space as ${\bf
C}^{(n|2m)}_\star$. The index $A_k$ takes the values $1,\,2$.
Other supermanifolds will be constructed as quotients of this
space. As a consequence we will then consider only supermanifolds
with an even number of fermionic dimensions. With this
modification the real submanifold $ z^i=e^{2i\theta_0^i}{\bar
z}^i,\,\psi^{A_k}=e^{2i\Theta_0^{A_k}}{\bar
 \psi}^{A_k}$ {\it is} Lagrangian since $d\psi_Ad\psi^A=0$. The new space ${\bf
C}^{(n|2m)}_\star$ is still, obviously, super Calabi-Yau. One
possible way to verify this claim is to check that the super
Monge-Ampere equation $\mbox {sdet}K_{i\bar{j}}=1$ is satisfied:
 \be
\mbox
{sdet}\left(%
\begin{array}{cccccc}
  \bf{1}_{n\times n}&  &  &  & & \\
   & 0 & 1 &  & &\\
   & -1 & 0 &  & & \\
 &  &  &  \ddots& &\\
&  &  & &0 &1\\
&  &  & & -1&0
\end{array}%
\right)=1.\label{risc} \ee
 In eq. (\ref{risc}) we used the definition of superdeterminant or
 Berezinian:
\begin{equation}
\mbox{sdet}\left(%
\begin{array}{cc}
  A & B\\
  C & D \\
\end{array}%
\right)={\mbox{det}(A-BD^{-1}C)\over{\mbox{det} D}}
\end{equation}
where $A,D$ and $B,C$ are respectively Grassmann even and
Grassmann odd matrices. We can now proceed in parallel with
bosonic case considering the equation
\begin{equation}
q_i^{\alpha}|z^i|^2+p_k^{\alpha}\epsilon_{A_k\,B_k}\psi^{A_k}\bar{\psi}^{B_k}=
c^{\alpha},\,\,\,\alpha=1,...,n-r.\label{supo}
\end{equation}
 We can explicitly solve eq.(\ref{supo})
for the bosonic variables ${|z^i|^2}$ as
\begin{equation}
|z^i|^2=v^i_{\beta}s^{\beta}-r^i_k
\epsilon_{A_k\,B_k}\psi^{A_k}\bar{\psi}^{B_k}+d^i\label{gen}
\end{equation}
with the following conditions
\begin{equation}
q_i^{\alpha}v^i_{\beta}=0,\,\,\,\,q_i^{\alpha}d^i_{\beta}=c^{\alpha},\,\,\,\,
q_i^{\alpha}r^i_k=p^{\alpha}_k.
\end{equation}
The locus has real superdimension $(n-(n-r))-2m=r-2m$. Using
eq.(\ref{supo}), the bosonic part of the super K\"{a}hler form
gives \be\label{gen1}
 d|z^i|^2\wedge
 d\theta^i=ds^{\beta}\wedge d(
 v^i_{\beta}\theta^i)-r^i_k\epsilon_{A_k\,B_k}
 (d\psi^{A_k}\bar{\psi}^{B_k}+\psi^{A_k}d\bar{\psi}^{B_k})\wedge d\theta^i.
\ee Using $\psi^{A_k}=e^{2i\Theta^k}{\bar
 \psi}^{A_k}$ and parameterizing the bosonic angles as
$\theta^i=q^i_{\al}\phi^{\alpha}$ this becomes
  \be\label{gen2}
 -e^{2\Theta^k}\epsilon_{A_k\,B_k}(2id\Theta^k\bar{\psi}^{A_k}\bar{\psi}^{B_k}
 +2d\bar{\psi}^{A_k}\,\bar{\psi}^{B_k})\wedge
 d(r^i_k\theta^i).\label{inc2}
\ee
 The fermionic part of the K\"{a}hler form reads instead
\begin{eqnarray}&&
ie^{2\Theta^k}\epsilon_{A_k\,B_k}(d{\bar
 \psi}^{A_k}d\bar{\psi}^{B_k}+2id\Theta^k\bar{\psi}^{A_k}d\bar{\psi}^{B_k})\nonumber\\
 &&=\,\,\,\, -2e^{2\Theta^k}
 d\Theta^k\bar{\psi}^{A_k}d\bar{\psi}^{B_k}\label{inc1}
\end{eqnarray}
where we used the property that the $d{\bar{\psi}}^{A_k}$'s
commute. The sum of (\ref{inc2}) and (\ref{inc1}) is zero if we
choose $r^i_k\theta^i=\Theta^k$. The Lagrangian is then a
$T^{n-r}$ fibration parametrized by $\{\phi^{\alpha}\}$ over the
locus (\ref{supo}), with
$\theta^i=q_i^{\alpha}\phi^{\alpha},\,\,\Theta^k=p_k^{\alpha}\phi^{\alpha}$.

The moment map associated to the $U(1)$ vector field \be
X=Q^iz^i{\partial\over \partial z^i}-Q^i\bar{z}^i{\partial\over
\partial \bar{z}^i}+P^k\psi^{A_k}{\partial\over \partial \psi^{A_k}}-P^{k}
\bar{\psi}^{A_k}{\partial\over \partial \bar{\psi}^{A_k}}
 \ee is
\be Q^i|z^i|^2+P^k\epsilon_{A_k\,B_k}{
 \psi}^{A_k}\bar{\psi}^{B_k}=r
\ee Note that to preserve the K\"{a}hler (\ref{kahler}) form we
have assigned the same charge $P^{k}$ to each couple of fermionic
fields $\psi^{A_k}$. The quotient ${\bf C}^{(n|2m)}_\star//U(1)$
then is a super Calabi-Yau iff\footnote{More generally if we have
the K\"{a}hler quotient ${\bf C}^{(n|2m)}_\star//U(1)^r$ the CY
condition is $\sum_{i=1}^{n}
Q^i_{\alpha}=2\sum_{k=1}^{m}P^k_{\alpha}$ with $\alpha=1,...,r.$ }
$\sum_{i=1}^{n} Q^i=2\sum_{k=1}^{m}P^k$.
 If we want the Lagrangian to descend to the Calabi-Yau quotient we
need to preserve the constraints $v^i\theta^i=0$ and
$r^i_k\theta^i=\Theta^k$. The action of the $U(1)$ group on the
phases is $\theta^i\rightarrow \theta^i+Q^i_{\al}\varphi^{\al}$
and $\Theta^k\rightarrow \Theta^k+P^k_{\al}\varphi^{\al}$ and
therefore we need \be
 v^iQ^i=0,\,\,\,\,\,r^i_kQ^i=P^k.\label{cond}
 \ee
 The special Lagrangian condition for the submanifold (\ref{supo}) is
 \be
\sum_{i=1}^{n} q_i^{\alpha}-2\sum_{k=1}^{m}p_k^{\alpha}=0.
 \ee
Let us consider some examples. We begin with
\begin{eqnarray}
|z^1|^2+|z^3|^2+\epsilon_{A_1\,B_1}\psi^{A_1}\bar{\psi}^{B_1}&=&\nonumber
c^{1}\\\label{slag0}
|z^2|^2+|z^4|^2+\epsilon_{A_2\,B_2}\psi^{A_2}\bar{\psi}^{B_2}&=&
c^{2} \end{eqnarray} in ${\bf C}^{(4|4)}_{\star}$. Note that the
special Lagrangian condition is satisfied. Performing a K\"{a}hler
quotient with charges $Q^i=1,\,i=1,...,4$ and $P^k=1,\,k=1,2$ we
obtain the super Calabi-Yau ${\bf CP}^{(3|4)}_{\star}$. One can
verify that the submanifold (\ref{slag0}) satisfies the conditions
(\ref{cond}) and therefore descends to a special Lagrangian in
${\bf CP}^{(3|4)}_{\star}$. As a further example we can take
\begin{eqnarray}
2|z^1|^2-|z^2|^2-|z^4|^2&=& c^{1}\\\nonumber\label{slag2}
|z^2|^2+|z^3|^2+\epsilon_{AB}\psi^{A}\bar{\psi}^{B}&=& c^{2}.
\end{eqnarray}
in the superprojective space ${\bf WCP}(-2,1,2,1|1,1)$ which is
obtained from ${\bf C}^{(4|2)}_\star$ dividing by the $U(1)_{\bf
C}$ group with charges $(Q^i|P^k)=(-2,1,2,1|1,1)$.

Modding out by the complexified gauge group $U(1)_{\bf C}$ always
reduces the complex bosonic dimension by one, without changing the
fermionic dimension. Since we cannot gauge away fermions we cannot
have submanifolds of the form $p_k
\epsilon_{A_k,B_k}\psi^{A_k}\bar{\psi}^{B_k}=c$. Therefore one
additional constraint comes from requiring that, when  considering
the matrix of the charges
\begin{displaymath}
\left(\begin{array}{c|c} Q^i& P^k\\
q_i^{\alpha} & p_k^{\alpha}
\end{array}\right),
\end{displaymath}
the bosonic submatrix
 \be \left(\begin{array}{c}
 Q^1,...\,, Q^n \\
  q_1^{\alpha},...\,, q_n^{\alpha}\\
\end{array}\right)
\ee has maximum  rank.

Let us now discuss how the special Lagrangian (\ref{supo}) map in
the dual Landau-Ginzburg theory. The only novelty comes from the
modified K\"{a}hler form for the fermionic directions. To learn
how to proceed let us study the following bosonic gauged linear
sigma model \be {\cal L}=\int
d^4\theta\,\left(i\epsilon_{AB}\bar{\Phi}^Ae^{2QV}\Phi^B -{1\over
2e^2}\bar{\Sigma}\Sigma\right)-{1\over 2}\int
d^2\tilde{\theta}\,t\Sigma+c.c.,\,\,\,A=1,2. \label{pal}\ee It is
convenient to make the field transformation \begin{eqnarray}
\Phi_1&=&\varphi_1+i\varphi_2\nonumber\\
\Phi_2&=&\varphi_2+i\varphi_1\label{transf}
 \end{eqnarray}
which enables to rewrite the kinetic term for the chiral fields as
$
-2(\bar{\varphi}_1e^{2QV}\varphi_1-\bar{\varphi}_2e^{2QV}\varphi_2)
$. We now introduce the following Lagrangian:
 \begin{eqnarray} {\cal L}&=&\int
d^4\theta\,\left(e^{2QV+B_1}-{1\over 2}(Y_1+\bar{Y}_1)B_1\right)
-\int d^4\theta\,\left(e^{2QV+B_2}-{1\over
2}(Y_2+\bar{Y}_2)B_2\right)\nonumber\\&&-\int d^4\theta{1\over
2e^2}\bar{\Sigma}\Sigma-{1\over 2}\int
d^2\tilde{\theta}\,t\Sigma+c.c..\label{L1}
 \end{eqnarray}
 The equations of motion of $Y_1$ and $Y_2$ imply that
 \be
B_1=\psi_1+\bar{\psi}_1,\,\,\,B_2=\psi_2+\bar{\psi}_2\label{prim}
 \ee
where $\psi_1$ and ${\psi}_2$ are two chiral fields. We obtain the
desired Lagrangian with the identification $\varphi_1=e^{\psi_1}$
and $\varphi_2=e^{\psi_2}$. Proceeding differently and integrating
out the $B$ fields gives
 \be
B_1=-2QV+Log[-{i\over
2}(Y_1+\bar{Y}_1)],\,\,\,B_2=-2QV+Log[-{i\over
2}(Y_2+\bar{Y}_2)]\label{sec}
 \ee
Inserting this expression in the enlarged Lagrangian we can read
off the classical dual twisted superpotential \be \tilde
W_{cl.}=\int d^2\tilde{\theta}Q\Sigma\left(Y_1-Y_2-t\right) \ee to
which one must add the instanton correction $\tilde
W_{inst.}=e^{-Y_1}-e^{-Y_2}$. By integrating out $\Sigma$ we
obtain ``the dual D-term condition'' $Y_1-Y_2=t$.
 The relation between the lowest components of the chiral fields
$\varphi_{A}$ and the dual twisted fields $Y_A$ is as usual \be
{1\over 2}\mbox{Re}Y_i= |\varphi_i|^2. \ee These considerations
suggest that, in the fermionic generalization and after having
done a field transformation similar to (\ref{transf}), the
equation
\begin{equation}
q_i^{\alpha}|z^i|^2+p_k^{\alpha}(|\psi_1^k|^2-|\psi_2^k|^2)=
c^{\alpha}
\end{equation} becomes in the dual variables
\begin{equation}
q_i^{\alpha}Y^i-p_k^{\al}(X^k_1-X^k_2)= c^{\alpha}.
\end{equation}
The dual Landau-Ginzburg superpotential is \be \tilde
W=\sum_{i=1}^n
e^{-Y^i}+\sum_{k=1}^me^{-X_1^k}(1+\eta_1^k\chi_1^k)-e^{-X_2^k}(1+\eta_2^k\chi_2^k)
\ee with D-term constraint \be \sum_{i=1}^n
Q_iY^i+\sum_{k=1}^mQ_k(X_1^k-X_2^k)=t. \ee

\section{Infinitesimal Deformations}
In this final section we want to comment on the space of
infinitesimal deformations of special Lagrangians inside a
supermanifold. Let us begin by reviewing the bosonic case.  There
is a quite convenient way to study the local geometry of a
Lagrangian in ${\bf C}^n$ which is familiar in symplectic
geometry\cite{joyce}. Locally every Lagrangian can be thought as
the graph $\Gamma_f$ of a closed 1 form $df$, where $f$ is a
smooth function from ${\bf R}^n$ to ${\bf R}$. This simply means
that the Lagrangian can be seen locally as the real
$n$-dimensional submanifold
\begin{equation}
\Gamma_f=\{(x^1,y^1=\partial_{x^1}f(x^1,...,x^n),...,x^{n},y^n=\partial_{x^n}f(x^1,...,x^n))
; x^1,...,x^n\in {\bf R}\}\label{rt}
\end{equation}
in ${\bf C}^n$. Indeed the restriction of the K\"{a}hler form is
$k_{\Gamma_f}=\partial^2_{i,j}f \,dx^i\wedge dx^j=0$. We would
like now to understand how to impose the special Lagrangian
condition in this formalism. Under the change of variables \be
{z^i}\rightarrow {z^i=x^i+i\partial_i f(x^1,...,x^n)}\ee we obtain
the following transformation rule for the holomorphic top form:
\begin{equation}
\prod_{i=1}^n dz^i= J \prod_{i=1}^n dx^i
\end{equation}
where the Jacobian $J$ is $\mbox{det}(I+i\mbox{Hess} f)$. Since
$\prod_i dx^i$ is real by construction, the special Lagrangian
condition, $\mbox{Im}\Omega_{|L}=0$, is then equivalent to

\begin{equation}
\mbox{Im\,det}(I+i\mbox{Hess} f)=0. \label{imdet}
\end{equation}
We can now study infinitesimal deformations of special Lagrangians
in ${\bf C}^n$. Using the fact that every Lagrangian looks locally
like ${\bf R}^n$ we can study the infinitesimal deformations of
${\bf R}^n$ which preserve the special Lagrangian condition. The
deformation of ${\bf R}^n$ can be seen as the graph $\Gamma_f$,
with the condition that the function $f$ and its derivatives are
infinitesimal. We can then linearize equation (\ref{imdet}) to
obtain
\begin{equation}
\mbox{Im\,det}(I+i\mbox{Hess} f)\sim \mbox{Tr\,Hess}=\triangle
f=0. \label{imdet2}
\end{equation}
 This result
shows that infinitesimal deformations of special Lagrangian in
${\bf C}^n$ are associated to harmonic functions on ${\bf R}^n$.
Since adding a constant to $f$ does not change $\Gamma_f$, the
submanifold (\ref{rt}) is parametrized by $df$. Infinitesimal
deformations of a special Lagrangian $\cal{L}$ correspond
therefore to harmonic 1-forms on $\cal{L}$. This result is a first
step toward the Mclean's theorem\cite{mclean} according to which
the moduli space of special Lagrangian deformations of a compact
Lagrangian $L$ is a smooth manifold of dimension $b^1(L)$.

We can now discuss the extension to the super Lagrangian case. We
consider for simplicity ${\bf C}^{(n|2)}_\star$. Using the
decomposition $z^i=x^i+iy^i$, $\,\psi^{A}=\eta^{A}+i\chi^{A}$, the
K\"{a}hler form
$\sum_{i=1}^{n}idz^id\bar{z}^i+\sum_{k=1}^{m}i\epsilon_{A\,B}d\psi^{A}d\bar{\psi}^{B}$
becomes \be
\omega=2\sum_{i=1}^{n}dx^idy^i+2\sum_{k=1}^{m}d\eta_{A}d\chi^{A}.
\ee The natural generalization of (\ref{rt}) is
\begin{equation}
\Gamma=\{z^i=x^i+i\partial_{x^i}f({x},{\eta}),\,\psi^{A}=\eta^{A}+ig^{A}({x},{\eta})
\}\label{graph}
\end{equation}
The restriction of the K\"{a}hler on this locus turns out to be
\begin{equation}
 2{\partial^2 f\over{\partial x^m\partial x^n}}dx^m\wedge
 dx^n+2 dx^m d\eta^{A}\left({\partial^2 f\over{\partial x^m\partial \eta^A}}+{\partial g_{A}
 \over{\partial x^m}}\right)
+2d\eta^{A} d\eta^{B} {\partial g_A\over{\partial
\eta^{B}}}.\end{equation} Requiring $k_{\Gamma}=0$ yields
\begin{equation}
g_{A}=-{\partial f\over\partial \eta^{A}},\,\,\,\,\,\,\,{\partial
g^{A}\over\partial \eta^{B}}=\delta^{A}_{B} h(x).
\end{equation} These
conditions imply that $g^{A}=\eta^{A} h(x)$ and $f=f_0(x)-{1\over
2}\eta^{A}\eta_{A} h(x)$. The top holomorphic form is
\begin{equation}
\prod_{i,{A}}dz^id\psi^{{A}}={\cal J}\prod_{i,{A}}dx^id\eta^{{A}}
\end{equation}where ${\cal J}$ is the super-Jacobian

\be {\cal J}=\mbox{sdet}\left(%
\begin{array}{cc}
  1+i{\mbox Hess}f & -i{\partial^2 f/{\partial x^m\partial \eta_{A}}} \\
  i{\partial^2 f/{\partial x^m\partial \eta^{A}}} & \delta^{A}_{B}(1+ih) \\
\end{array}%
\right)\ee

To study local deformations we specialize to the Lagrangian
$z^i=e^{2i\theta^i_0}{\bar z}^i,\, \psi^A=e^{2i\Theta^{A}_0}{\bar
\psi}^{A}$ in ${\bf C}^{(n|2)}_\star$. A Lagrangian which differs
from this one by an infinitesimal deformation looks then locally
like (\ref{graph}), with the condition that $f$ and its
derivatives are kept small. To require that the deformation is
special Lagrangian we need to impose $\mbox{Im}{\cal J}=0$ which,
to linear order in the deformation, is equivalent to
 \be
 \mbox{Im}\,{\mbox{det}(1+i\mbox{Hess}f)\over \mbox{det}[\delta^{A}_{B}(1+ih)]}\sim\triangle f-h=0, \ee where, as
before, $\triangle$ is the ordinary Laplacian in ${\bf R}^n$. The
last equation splits into \be
 \triangle
f_0=h,\,\,\,\,\triangle h=0. \ee This suggests that special
Lagrangian deformations are associated to a pair of harmonic
functions $h$ and $f_0^h$, the second being a solution of the
homogeneous equation for $f_0$. Extrapolating this result we would
expect a moduli space of dimension $b_1(L)^2$ for compact special
Lagrangians. One can easily extend this result to Lagrangian
submanifolds in ${\bf C}^{(n|m)}_\star$.

\section{Acknowledgement}
It is a pleasure to thank Daniel Robles-Llana for useful comments
at an early stage of the project and Martin Ro\v{c}ek for
insightful discussions. I also acknowledge partial financial
support through the NSF award PHY-0354776.


\begin{thebibliography}{99}

\bibitem{witten}
E.~Witten, ``Perturbative gauge theory as a string theory in
twistor space,'' Commun.Math.Phys. 252 (2004) 189-258
[arXiv:hep-th/0312171].
\bibitem{Cachazo}{ F.~Cachazo, P.~Svrcek and E.~Witten,
``MHV vertices and tree amplitudes in gauge theory,'' JHEP {\bf
0409}, 006 (2004) [arXiv:hep-th/0403047];
}
{ G.~Georgiou, E.~W.~N.~Glover and V.~V.~Khoze, ``Non-MHV tree
amplitudes in gauge theory,'' JHEP {\bf 0407}, 048 (2004)
[arXiv:hep-th/0407027];
}
F.~Cachazo, P.~Svrcek and E.~Witten, ``Twistor space structure of
one-loop amplitudes in gauge theory,'' JHEP {\bf 0410}, 074 (2004)
[arXiv:hep-th/0406177];
A.~Brandhuber, B.~J.~Spence and G.~Travaglini, ``One-loop gauge
theory amplitudes in N = 4 super Yang-Mills from MHV
Nucl.\ Phys.\ B {\bf 706}, 150 (2005) [arXiv:hep-th/0407214].
\bibitem{Khoze:2004ba}
V.~V.~Khoze, ``Gauge theory amplitudes, scalar graphs and twistor
space,'' arXiv:hep-th/0408233;
\bibitem{Cachazoreview}
F.~Cachazo and P.~Svrcek, ``Lectures on twistor strings and
perturbative Yang-Mills theory,'' PoS {\bf RTN2005}, 004 (2005)
[arXiv:hep-th/0504194];
\bibitem{Berkovits}
N.~Berkovits and E.~Witten, ``Conformal supergravity in
twistor-string theory,'' JHEP {\bf 0408}, 009 (2004)
[arXiv:hep-th/0406051];
{ R.~Roiban, M.~Spradlin and A.~Volovich, ``A googly amplitude
from the B-model in twistor space,'' JHEP {\bf 0404}, 012 (2004)
[arXiv:hep-th/0402016];
}
P.~A.~Grassi and G.~Policastro, ``Super-Chern-Simons theory as
superstring theory,'' arXiv:hep-th/0412272.
\bibitem{Neitzke}
A.~Neitzke and C.~Vafa, ``N = 2 strings and the twistorial
Calabi-Yau,'' arXiv:hep-th/0402128;


\bibitem{gravity}
S.~Giombi, R.~Ricci, D.~Robles-Llana and D.~Trancanelli, ``A note
on twistor gravity amplitudes,'' JHEP {\bf 0407}, 059 (2004)
[arXiv:hep-th/0405086];
Z.~Bern, N.~E.~J.~Bjerrum-Bohr and D.~C.~Dunbar, ``Inherited
twistor-space structure of gravity loop amplitudes,'' JHEP {\bf
0505}, 056 (2005) [arXiv:hep-th/0501137];
 V.~P.~Nair,
``A note on MHV amplitudes for gravitons,'' Phys.\ Rev.\ D {\bf
71}, 121701 (2005) [arXiv:hep-th/0501143];
J.~Bedford, A.~Brandhuber, B.~J.~Spence and G.~Travaglini, ``A
recursion relation for gravity amplitudes,'' Nucl.\ Phys.\ B {\bf
721}, 98 (2005) [arXiv:hep-th/0502146];
F.~Cachazo and P.~Svrcek, ``Tree level recursion relations in
general relativity,'' arXiv:hep-th/0502160.
\bibitem{rey}
M.~Kulaxizi and K.~Zoubos, ``Marginal deformations of N = 4 SYM
from open / closed twistor strings,'' arXiv:hep-th/0410122.
{ J.~Park and S.~J.~Rey, ``Supertwistor orbifolds: Gauge theory
amplitudes and topological strings,'' JHEP {\bf 0412}, 017 (2004)
[arXiv:hep-th/0411123].
}; S.~Giombi, M.~Kulaxizi, R.~Ricci, D.~Robles-Llana,
D.~Trancanelli and K.~Zoubos, ``Orbifolding the twistor string,''
Nucl.\ Phys.\ B {\bf 719}, 234 (2005) [arXiv:hep-th/0411171];

\bibitem{PopovRB}
  A.~D.~Popov and C.~Saemann,
   ``On supertwistors, the Penrose-Ward transform and N = 4 super Yang-Mills
  theory,''
  Adv.\ Theor.\ Math.\ Phys.\  {\bf 9}, 931 (2005)
  [arXiv:hep-th/0405123].
  A.~D.~Popov and M.~Wolf,
   ``Topological B-model on weighted projective spaces and self-dual models  in
  four dimensions,''
  JHEP {\bf 0409}, 007 (2004)
  [arXiv:hep-th/0406224].
C.~Saemann, ``The topological B-model on fattened complex
manifolds and subsectors of N = 4 self-dual Yang-Mills theory,''
JHEP {\bf 0501}, 042 (2005) [arXiv:hep-th/0410292].
  M.~Wolf,
  ``On hidden symmetries of a super gauge theory and twistor string theory,''
  JHEP {\bf 0502}, 018 (2005)
  [arXiv:hep-th/0412163].
  A.~D.~Popov, C.~Saemann and M.~Wolf,
   ``The topological B-model on a mini-supertwistor space and supersymmetric
  Bogomolny monopole equations,''
  JHEP {\bf 0510}, 058 (2005)
  [arXiv:hep-th/0505161].
  C.~Saemann,
  ``On the mini-superambitwistor space and N = 8 super Yang-Mills theory,''
  arXiv:hep-th/0508137.




\bibitem{ma1}
 M.~Rocek and N.~Wadhwa, ``On Calabi-Yau supermanifolds,''
[arXiv:hep-th/0408188]
{ M.~Rocek and N.~Wadhwa, ``On Calabi-Yau supermanifolds. II,''
[arXiv:hep-th/0410081]
};
{ C.~g.~Zhou, ``On Ricci flat supermanifolds,'' JHEP {\bf 0502},
004 (2005) [arXiv:hep-th/0410047]
};
\bibitem{Lindstrom:2005uh} U.~Lindstrom, M.~Rocek and R.~von Unge,
``Ricci-flat supertwistor spaces,'' arXiv:hep-th/0509211.

\bibitem{Witten:1978xx}
E.~Witten,
Phys.\ Lett.\ B {\bf 77}, 394 (1978).

\bibitem{Sethi}
S.~Sethi, ``Supermanifolds, rigid manifolds and mirror symmetry,''
Nucl.\ Phys.\ B {\bf 430}, 31 (1994) [arXiv:hep-th/9404186].
\bibitem{Aganagic:2004yh}
M.~Aganagic and C.~Vafa, ``Mirror symmetry and supermanifolds,''
arXiv:hep-th/0403192.
\bibitem{harvey}
F.R. Harvey and H.B. Lawson, ``Calibrated Geometries,'' Acta Math.
(1982),47-157
\bibitem{Kumar}
S.~P.~Kumar and G.~Policastro, ``Strings in twistor superspace and
mirror symmetry,'' Phys.\ Lett.\ B {\bf 619}, 163 (2005)
[arXiv:hep-th/0405236].
\bibitem{ahn}{ C.~h.~Ahn,
``Mirror symmetry of Calabi-Yau supermanifolds,'' Mod.\ Phys.\
Lett.\ A {\bf 20}, 407 (2005) [arXiv:hep-th/0407009].
}
\bibitem{belhaj}
A.~Belhaj, L.~B.~Drissi, J.~Rasmussen, E.~H.~Saidi and A.~Sebbar,
``Toric Calabi-Yau supermanifolds and mirror symmetry,'' J.\
Phys.\ A {\bf 38}, 6405 (2005) [arXiv:hep-th/0410291].
\bibitem{Schwarz:1995ak}
A.~Schwarz, ``Sigma models having supermanifolds as target
spaces,'' Lett.\ Math.\ Phys.\  {\bf 38}, 91 (1996)
[arXiv:hep-th/9506070].
\bibitem{aga}
M.~Aganagic and C.~Vafa, ``Mirror symmetry, D-branes and counting
holomorphic discs,'' arXiv:hep-th/0012041.
\bibitem{klemm}
M.~Aganagic, A.~Klemm and C.~Vafa, ``Disk instantons, mirror
symmetry and the duality web,'' Z.\ Naturforsch.\ A {\bf 57}, 1
(2002) [arXiv:hep-th/0105045].

\bibitem{hori}{ K.~Hori and C.~Vafa,
``Mirror symmetry,'' arXiv:hep-th/0002222.
}

\bibitem{phase}{ E.~Witten,
``Phases of N = 2 theories in two dimensions,'' Nucl.\ Phys.\ B
{\bf 403}, 159 (1993) [arXiv:hep-th/9301042].
}
\bibitem{gia}{ S.~Seki and K.~Sugiyama,
``Gauged linear sigma model on supermanifold,''
arXiv:hep-th/0503074.
}


\bibitem{joyce}
D.~Joyce,\,``Lectures on special Lagrangian geometry,''
[arXiv:math.dg/0111111].
\bibitem{mclean}
R.C. McLean, ``Deformations of calibrated submanifolds,''
Communications in Analysis and Geometry {\bf}6 (1998),705-747.


\end{thebibliography}
\end{document}